\documentclass[showpacs,aps,prappled,twocolumn,superscriptaddress,10pt,longbibliography]{revtex4-1}
\usepackage{graphicx}

\usepackage{bm}
\usepackage{color}
\usepackage[colorlinks=true, allcolors=blue]{hyperref}
\usepackage{enumerate}
\usepackage{xspace}
\usepackage{mathrsfs}
\usepackage{mathptmx}
\usepackage[utf8]{inputenc}
\usepackage{enumitem}
\usepackage{amssymb,amsmath}

\usepackage{graphicx}
\usepackage{listings}
\usepackage{float}
\usepackage{color}
\usepackage{multirow}
\usepackage{multirow}
\usepackage[table]{xcolor}
\newcommand{\vampire}{\textsc{vampire} }
\newcommand{\Tc}{\ensuremath{T_{\mathrm{C}}}\xspace}

\begin{document}

\title{Model of damping and anisotropy at elevated temperatures: application to granular {FePt} films }
\author{Mara Strungaru}
\author{Sergiu Ruta}
\author{Richard~F.~L.~Evans}
\author{Roy W. Chantrell} 
\affiliation{Department of Physics, University of York, York, YO10 5DD, UK}

\begin{abstract}
 Understanding the damping mechanism in finite size systems and its dependence on temperature is a critical step in the development of magnetic nanotechnologies. In this work, nano-sized materials are modeled via atomistic spin dynamics, the damping parameter being extracted from Ferromagnetic Resonance (FMR) simulations applied for FePt systems, generally used for heat-assisted magnetic recording media (HAMR). We find that the damping increases rapidly close to $T_{\mathrm{C}}$ and the effect is enhanced with decreasing system size, which is ascribed to scattering at the grain boundaries. Additionally, FMR methods provide the temperature dependence of both damping and the anisotropy, important for the development of HAMR. Semi-analytical calculations show that, in the presence of a grain size distribution, the FMR linewidth can decrease close to the Curie temperature due to a loss of inhomogeneous line broadening. Although FePt has been used in this study, the results presented in the current work are general and valid for any ferromagnetic material.
\end{abstract}

\maketitle
\section{Introduction}

The magnetic damping parameter is important from both a fundamental and applications point of  view  as  it controls the dynamic properties of the system such as magnetic relaxation, spin waves, domain wall propagation and magnetic reversal processes. Magnetic materials have a broad range of interest for nano-devices/nano-elements and exhibit a fast response to external excitations. In information technologies, damping plays a crucial role, especially for spin-transfer torque magnetic random access memories (STT-MRAM) where it controls the switching current \cite{slonczewski1996current}.  With the emerging field of magnetisation switching via ultrafast laser pulses, the damping parameter can influence the fluence of the laser pulse necessary for demagnetising and switching of the sample \cite{atxitia2015optimal}. Spintronic devices such as race-track memories which are based on domain wall propagation in magnetic nanowires are also influenced by damping \cite{yuan2016influence}. As current  magnetic technologies are based on nanostructures of smaller and smaller sizes, the finite size effects become more important and can significantly influence the magnetic properties including the damping. Therefore understanding the dependence of damping on temperature in finite size systems is a critical step in the development of magnetic nano-technologies.

One of the technologies that is strongly influenced by damping is magnetic recording, where the damping constant of the storage medium controls the writing speeds and bit error rates \cite{lyberatos2003thermal,kobayashi2017impact}. The next generation of ultra-high density storage technology is likely to be based on heat-assisted magnetic recording (HAMR) \cite{Rottmayer2006, Weller2014, Weller2016, kryder2008heat}. The main candidate for HAMR media is $L1_0$ ordered FePt \cite{Weller2016, Hu2013} due to its large perpendicular anisotropy and low Curie temperature (\Tc). For HAMR applications, information is stored at room temperature (300K) but the writing is done at elevated temperatures close to \Tc, therefore providing a large  range over which the temperature dependence of the damping needs to be understood. As the areal density increases, the grain size decreases and finite-size effects are becoming crucial. For this reason, FePt is an ideal candidate for studying  temperature and finite size effects on the damping. Although FePt has been used in this study, the  results presented in the current work are general and valid for any ferromagnetic material.

First investigations of the Gilbert damping for FePt 
involved experimental measurements via optical pump-probe techniques. The damping measured at room temperature  varies widely from one study to another. Becker et al. \cite{Becker2014} reported an effective damping of 0.1, and an even larger value (0.21) was found by Lee et al. \cite{Kyeong-Dong2014}, while the measurments of Mizukami et al \cite{Mizukami2011} gave a value of 0.055. It is important to note that these values include both intrinsic and extrinsic contributions, the purely intrinsic damping being even smaller than the reported values  \cite{Mizukami2011,mo2008origins}. Recently, Richardson et al \cite{Richardson2018} reported experimental measurements of damping at elevated temperatures showing an unexpected decrease of damping with temperature. A decrease in the effective damping can be crucial in HAMR, as this can increase the switching time,  affect the signal to noise ratio  and negatively impact the performance of HAMR. Theoretical studies on how the damping varies at elevated temperatures and in finite sized systems is therefore a critically important problem.

Ostler \textit{et al} \cite{Ostler2014} have successfully calculated the temperature dependence of damping in FePt bulk and thin-film systems based on the Landau-Lifshitz-Bloch (LLB) equation~\cite{garanin1997fokker}, showing an increased damping for thin-film systems, in comparison with the bulk case. The LLB equation is derived for a bulk material. It is important to note that a major contribution to damping, especially at elevated temperatures, arises from magnon scattering. On the bulk scale these processes are reproduced by the LLB equation, but with decreasing linear dimension, finite size and surface effects become important. Since these are not accounted for by the LLB equation it is necessary to use atomistic spin dynamics (ASD) simulations \cite{Evans2014} for  nanoscale grains, as ASD calculations include magnon processes. Using atomistic spin dynamics, ASD, we are able to calculate the FMR spectra  for small system sizes at elevated temperatures. Ferromagnetic resonance  simulations are computationally very expensive, hence we have developed a more efficient method of calculating the damping of these systems via a grid search method. We show that both of the methods agree, and furthermore, that they are able to calculate both the dependence of damping and anisotropy as function of temperature. The dependence of anisotropy as function of temperature is crucial for HAMR as it defines the temperature at which the writing process occurs.

\section{Ferromagnetic resonance using atomistic spin dynamics}\label{section_FMR}
To calculate damping as function of temperature we perform atomistic spin dynamics (ASD) simulations using the software package \vampire \cite{Evans2014}. ASD simulations assume a fixed lattice of atoms to which is associated a magnetic  moment or spin $\mathbf{S}_i= \boldsymbol{ \mu }_i / \mu_s$ that can precess in an effective field $\mathbf{H}_{i}$ according to the Landau-Lifshitz Gilbert (LLG) equation.
In our model, the Hamiltonian of the system contains a Heisenberg exchange term of strength $J_{ij}$, uniaxial energy of strength $k_u$ and a Zeeman term as shown in Eq.~\ref{gen_ham}:

\begin{equation}
\mathcal{H} = -\frac{1}{2}\,  \sum_{i,j } J_{ij}\, (\mathbf{S}_i\, \cdot  \mathbf{S}_j\:)-k_u \sum_{i} (\mathbf{S}_i\, \cdot  \mathbf{e})^2 - \sum_{i} \mu_i  (\mathbf{S}_i\, \cdot  \mathbf{B})
\label{gen_ham}
\end{equation}
 The effective field can be calculated from the Hamiltonian of the model to which we add a thermal noise $ \xi_i$, that acts as a Langevin thermostat:

\begin{equation}
    \mathbf{H}_i= - \frac{1}{\mu_i \mu_0} \frac{\partial {\mathcal{H}} }{\partial \mathbf{S}_i} + \xi_i
\label{LLG2} 
\end{equation}

The thermal field is assumed to be a white noise, with the following mean, variance and strength, as calculated from the Fokker-Planck equation: 
\begin{equation}
 \langle \xi_{i\alpha}(t) \rangle=0, ~~  \langle \xi_{i\alpha}(t)\xi_{j\beta}(s) \rangle= 2D \delta_{\alpha,\beta} \delta_{ij} \delta(t-s)
\label{eff}
\end{equation}
\begin{equation}
D= \frac{ \lambda  k_B T }{\gamma \mu_i \mu_0}
\end{equation}
where $\lambda$  represents the coupling to the heat bath, $T$ the thermostat temperature, $\gamma$ the gyromagnetic ratio. We note that the heat bath coupling constant $\lambda$ is different from the effective Gilbert damping $\alpha$, as the latter includes contributions from magnon scattering and other extrinsic processes such as inhomogeneous line broadening.

After calculating the effective field that acts on each atom, the magnetisation dynamics is given by solving the LLG equation (Eq.~\ref{llg}) applied at the atomistic level \cite{EllisLTP2015} using a numerical integration based on the Heun scheme. 
\begin{equation}
 \frac{\partial \mathbf{S}_i }{\partial t}   = -\frac{\gamma}{(1+\lambda^2) } \mathbf{S}_i \times  (\mathbf{H}_{i}+   \lambda \mathbf{S}_i \times \mathbf{H}_{i} )   
 \label{llg}
\end{equation}

\begin{figure}[!tb]
\centering
  \includegraphics[clip, trim={3cm 8cm 0 0cm},width=0.9\linewidth]{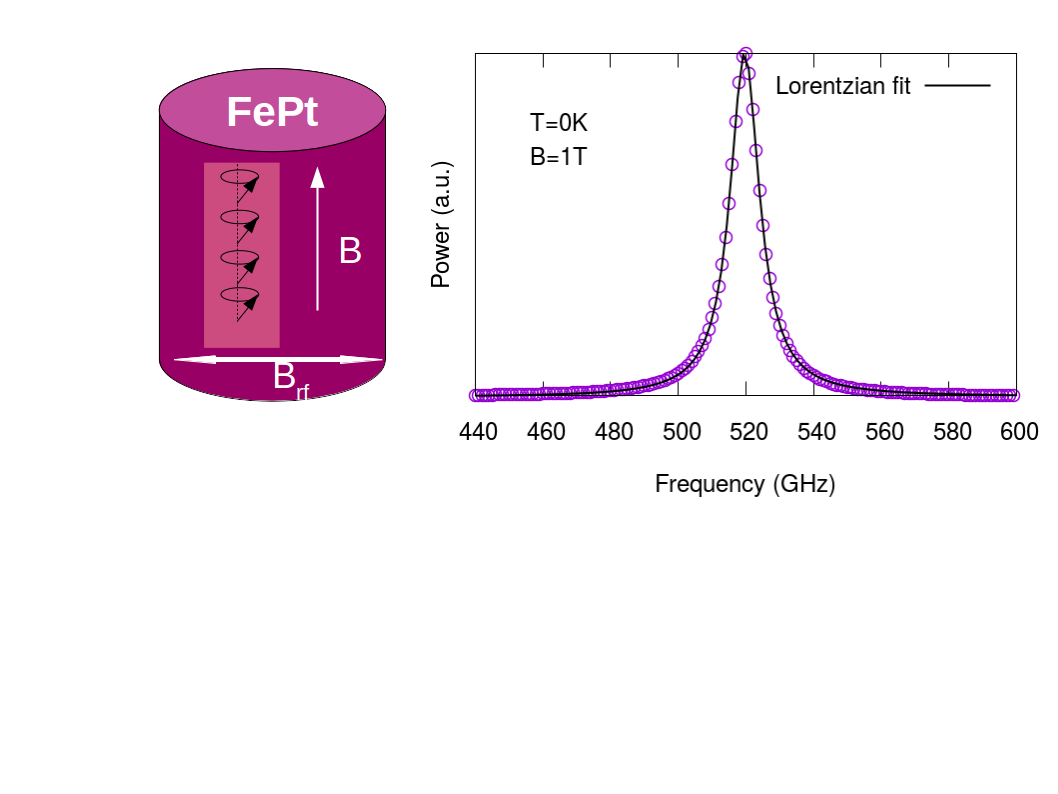}
  \caption{Illustration of the setup used for ferromagnetic resonance experiments. An out of plane magnetic field ($B$) and an in-plane oscillating field ($B_{rf}= B_0  \sin (2 \pi \nu t)$) is applied to the sample, as shown in the right inset.  By Fourier Transformation of the in-plane magnetisation the power spectrum as function of frequency is obtained. The simulation is performed for a single FePt spin at T=0K, having a damping of 0.01. This is equivalent with simulating a macrcospin at T=0K with equivalent properties. By fitting the power spectrum, the input resonance frequency and damping can be reproduced.  }\label{setup_example}
\end{figure}

To calculate the damping using atomistic spin dynamics we apply an out-of-plane magnetic field ($B$) to the sample with an additional in-plane oscillating field ($B_{\mathrm{rf}}= B_0  \sin (2 \pi \nu t)$), which is the default setup for ferromagnetic resonance experiments. The oscillating field will induce a coherent precession of the spins of the system which will result in an oscillatory behaviour of the in-plane magnetisation. By sweeping the frequency of the in-plane field, the  amplitude of the oscillations of magnetisation will change, with a maximum corresponding to the resonance frequency (as shown in Fig. \ref{setup_example}). By Fourier transformation of the in-plane magnetisation, the power spectrum as a function of frequency is obtained.  Fig. \ref{setup_example} shows the FMR spectra for a single spin of FePt at 0K. The spectra can be fitted by a Lorentzian curve (Eq.~\ref{lorenz}) where $w$ represents the width of the curve and A its amplitude. By fitting with Eq.~\ref{lorenz}, the effective Gilbert damping $\alpha$ and resonance frequency $f_0$ can be extracted.

\begin{equation}\label{lorenz}
L(x)= \frac {A} {\pi} \frac{0.5 w} {(x-f_0)^2+(0.5 w)^2}~, ~~\alpha= \frac {0.5 w }{ f_0}
\end{equation} 
 
The model parameters for FePt are listed in Tab.~1. L10 FePt has a face-centred tetragonal structure formed of alternating layers of Fe and Pt, which can be approximated to a  body-centered tetragonal structure with the central site occupied by Pt. The ab-initio calculations by Mryasov et al \cite{mryasov2005temperature} showed that the Pt spin moment is found to be linearly dependent on the exchange field from the neighbouring Fe moments. This dependence allows the Hamiltonian to be written only considering the Fe degrees of freedom. Under these assumptions, by neglecting the explicit Pt atoms, the system can be modelled as a simple cubic tetragonal structure with each atomic site corresponding to an effective Fe+Pt moment. The model used for the FePt system is restricted only to nearest neighbour interaction to minimise the computational cost of FMR calculations, in contrast with the full Hamiltonian given by Mryasov \textit{et al} \cite{mryasov2005temperature}. The nearest-neighbour exchange value is chosen to give a  Curie temperature of FePt of 720K, to be in agreement with reported values for nearest and long-range exchange magnetic Hamiltonian \cite{hovorka2012curie}. The damping parameter has been chosen to approximate the experimentally measured value in recording media provided by Advanced Storage Research Consortium (ASRC). The L$1_0$ phase of FePt has a very large uniaxial anisotropy, hence the increased thermal stability of the grains. The uniaxial anisotropy used in the simulation gives a anisotropy field of $H_k=2k_u/\mu_s=17.55$ T, slightly larger than the value used by Ostler et al (15.69 T). The FMR fields ($0.05$ T) used in our simulations are generally larger than experimental FMR fields to allow more accurate simulations with enhanced temperature. Our tests confirm that no non-linear modes are excited during the FMR simulations.

\begin{center}
\begin{table}[ht!]
\centering
\fontsize{8}{6}
\begin{tabular}{c|lll}
Quantity  & Symbol  & Value   &   Units   
 \\ \hline \hline 
Nearest-neighbours exchange & $J_{ij}$  & $6.71 \times 10^{-21}$    & J        \\
Anisotropy energy & $k_u$  & $2.63 \times 10^{-22}$    & J        \\
Magnetic moment & $\mu_S$  & $3.23$    & $\mu_B$       \\
Thermal bath coupling & $\lambda$  & $0.05$    &     \\
DC perpendicular field & $B$  & $1$    & T     \\
RF in-plane field & $B_{rf}$  & $0.05$    & T     \\ \hline
\end{tabular}
\\
\caption [Parameters used for FePt;]{Parameters used for the initial calculations of the damping constant of  FePt.}  
\end{table}
\end{center}

At $T = 0$K, the damping we extract from the FMR spectrum should correspond to the input coupling $\lambda$ as no thermal scattering effects are present, hence the effective damping of the system is given by the Gilbert damping which is the coupling to the heat bath. For this simulation, we have used an input heat bath constant of $\lambda=0.01$, which we then recover by performing FMR calculations at $T=0$K, method that serves as verification of our model.The damping obtained agrees within 0.1\% fitting error. The resonance peak should appear exactly at the resonance frequency given by Kittel formula $f_ {\mathrm{Kittel}}= \frac {\gamma}{2\pi} \cdot (B+\frac{2k_u}{\mu_S})$, depending on the applied field strength $(B)$ and on the perpendicular anisotropy of our system $(H_k=\frac{2k_u}{\mu_s})$. For an FePt system the resonance frequency we obtain is 520 GHz within 1\% fitting error, due to the exceptionally large magnetic anisotropy of the system. 

\section{Grid-search method}
\label{grid_search_section}
The Gilbert damping can be also calculated by fitting the time-traces of the magnetisation relaxation. The time-traces can be obtained via pump-probe experiments \cite{Becker2014}, however the dynamics of the magnetisation will include the effect of the laser pulse, such as heating and induced local magnetisations due to the inverse Faraday effect. To avoid the contributions to the damping from the laser pulse, damping can be calculated by taking the system out of equilibrium, letting it relax and subsequently recording the time-trace of the magnetisation. Ellis \textit{et al} \cite{Ellis2012} have numerically studied the damping of rare-earth doped permalloy using the transverse relaxation curves, by fitting them with the analytical solutions of the LLG equations. In the case of large anisotropy, exchange interaction and applied field, there is no simple general solution to the LLG equation. Pai \textit{et al} \cite{PaiPRA2019} used an applied field much larger than the anisotropy field so that the dynamics closely approximate that of the LLG equation with no anisotropy. However, this approach is unsuitable for FePt due to the very large fields required and also the influence of strong magnetic fields near the Curie temperature. Hence, we adopt a computational grid search method where we pre-calculate single spin solutions for the LLG equation using ASD, build a data base using these solutions and then build an algorithm that can identify the damping and anisotropy parameters from any transverse relaxation curve. The method we chose simply involves sweeping through the parameter space, the solution being given by minimising the sum of the squared residuals, a method known as grid-search. 

The grid search method can be used to fit time-dependent $m(t)$ curves in the case where analytical solutions do not exist. The numerical curves that need to be fitted are compared with each of the pre-calculated numerical curves with the single spin system. The best match will be given by the curve with lowest sum of squared residuals, the $\chi^2$ parameter, where $\chi^2$ is defined as:

\begin{equation}
\chi^2=\sum_{i=1}^N \Big[ {m(t^i)-f(t^i,\textbf{p})} \Big]  ^ 2
\end{equation}
where $m_i(t^i)$ is the value of the magnetisation at each moment in time $t^i$, $f(t^i,\textbf{p})$ are the pre-calculated single-spin dependences of the magnetisation at each moment $t^i$, $\textbf{p}$ is the list of parameters that have been varied (in our case $\textbf{p}$= ($K$, $\alpha$)). The minimum value of $\chi^2$ from all $\textbf{p}$ parameters is the best-agreement numerical solution.

\begin{figure}[!tb]
\centering
\includegraphics[trim={0 1cm 1cm 1cm},width=1.05\linewidth]{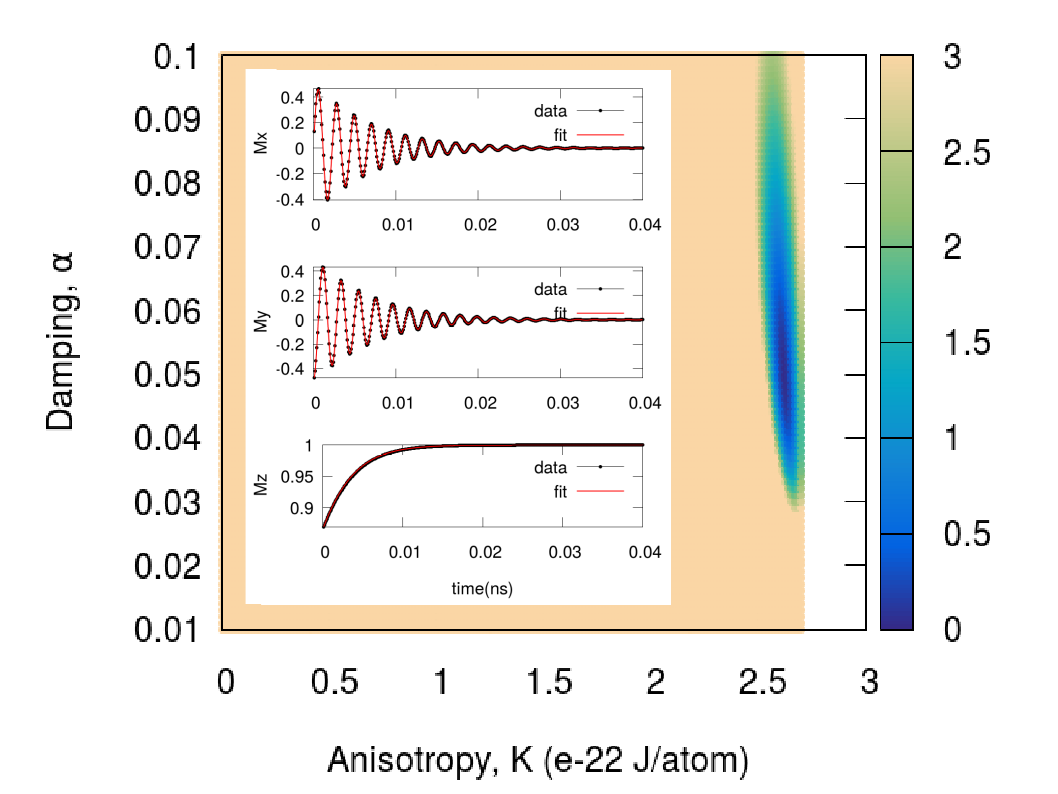}
\caption{ $\chi^2$ map calculated using the grid-search method based on single-spin simulations at T=0.1K. (inset) The input and fitted magnetisation relaxation curves showing the validation of the method. } 
\label{grid_search}
\end{figure}

Figure \ref{grid_search} shows the calculated $\chi^2$ as function of the main parameters, specifically the anisotropy and damping, at T=0.1K. In order to construct the single spin simulation data-base, we chose a resolution of $\Delta k_u=0.015 \times 10^{-22} J$ for the anisotropy and $\Delta \alpha_{\mathrm{step}}=0.001$  for the damping. It can be seen that the anisotropy is very well resolved: there is a sharp minimum at $k_u=2.625 \times 10^{-22}$, which is the closest value to the input anisotropy, $k_u=2.63 \times 10^{-22}$ taking into account the resolution we use for the data base. In the case of damping, the minimum is wider, leading to an error of approximately $ 0.017$ in determination of damping, which is slightly larger than the resolution used in the construction of the database.

\section{High-temperature FMR: damping and anisotropy calculations}

\begin{figure}[!tb]
\includegraphics[trim={0 0cm 0 0cm},clip,width=0.9\linewidth]{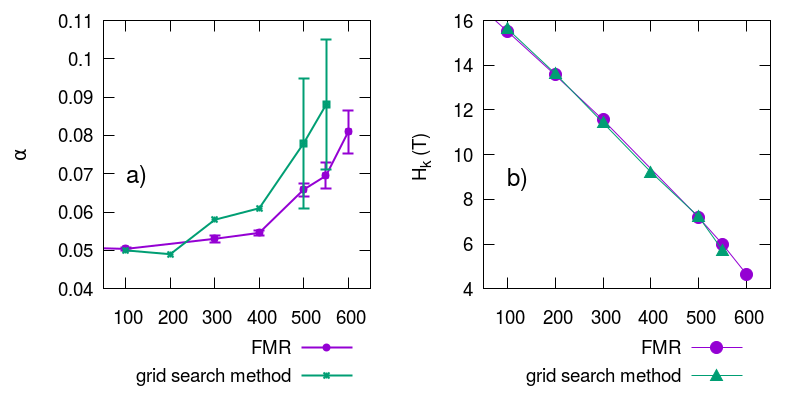} 
\caption{Comparison of FMR and grid-search fitting; a) Damping; b) Anisotropy; }
\label{comparison}
\end{figure}

In this section the damping and anisotropy are computed,  from frequency dependent FMR spectra and via the grid search method. The aim is to investigate the damping close to $T_c$ and in particular the effect of finite grain size. First, we test the effectiveness of the grid search method which was presented in Section \ref{grid_search_section}. Fig. \ref{comparison} shows the comparison between the two methods of calculation of the damping as a function of temperature for a granular system of 15 non-interacting grains of 5nm diameter and 10nm height.  The variations of anisotropy with temperature agree very well between the two methods, however the grid search method is far more computationally efficient. The enhanced computational efficiency comes from the fact that instead of simulating multiple frequency points to obtain the FMR, a single transverse relaxation simulation is needed to calculate the same parameters. The time-scales for the two simulations are also different: the frequency dependent FMR requires around 3 ns for each data point in the FMR spectra to perform the FFT analysis, while the transverse relaxation method requires, depending on the material, less than 1 ns. Extracted damping values agree reasonably well between the two methods, within the error bars. For the grid search method, there will be a damping interval that gives the same value of $\chi^2$. 
For the FMR experiment, the error bar is computed as the standard error of groups of 5 non-interacting grains. 

Because of the large error bars, especially close to \Tc for the grid search method,  we use the direct FMR simulations for the remainder of the paper and later consider possible means of improvement of the reliability of the grid search method.
For initial calculations we model a granular FePt system as a cylinder of 10nm height and 5nm diameter.
For comparison, the bulk FePt system is modelled via a system of $32 \times32\times32$ atoms with periodic boundary conditions.  Close to \Tc, the thermal fluctuations  become increasingly large for non-periodic systems and can lead to large errors in the determination of damping and anisotropy. For this reason, to reduce the statistical fluctuations, a system of 15 non-interacting grains is modelled. This significantly reduces the fluctuations in the magnetisation components and leads to statistically improved results. The in-plane magnetisation time series is Fourier transformed, and the damping is extracted as presented in Section \ref{section_FMR}.

\begin{figure}[!tb]
\centering
\includegraphics[width=0.9\linewidth]{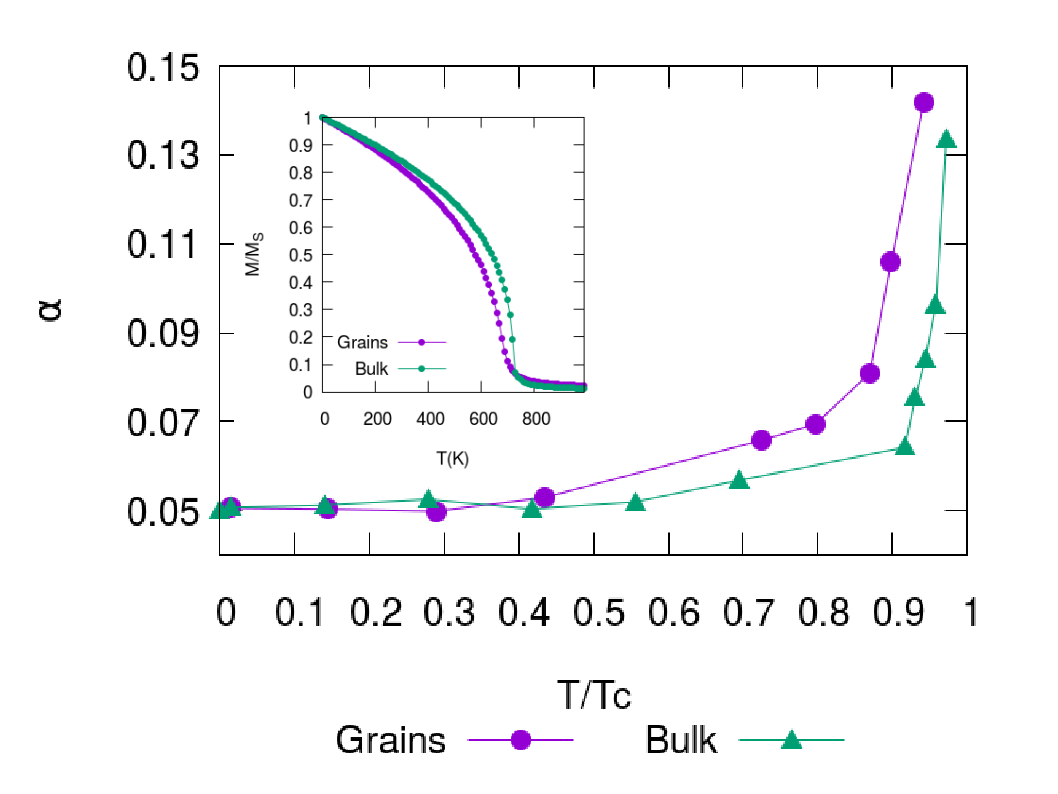}
\caption{Damping as function of normalised temperature for bulk and granular FePt system. The granular system shows overall larger damping than the bulk system, due to additional magnon scattering processes at the interface.  (inset) Magnetisation curves for granular and bulk FePt. The Curie temperatures for the two systems are : grain- $\Tc$=690K, bulk - $\Tc$ =720K.}
\label{fmr_damping_grain}
\end{figure}

Fig. \ref{fmr_damping_grain} shows the damping as a function of temperature for bulk and granular systems. For  comparison the temperature is normalised to the Curie Temperature of the systems, which differ due to finite size effects \cite{lyberatos2012size, hovorka2012curie}. The granular system will have a reduced Curie Temperature due to the cutoff in the exchange interactions at the surface. This is shown as an inset in Fig. \ref{fmr_damping_grain}, where the magnetisation as a function of temperature is computed  for the two systems. The Curie temperatures for the two systems, determined from the susceptibility peak, are: $\Tc$ for the grain =690K, and for the bulk $\Tc$ =720K. The input Gilbert damping parameter is 0.05, this value being reproduced at T=0K as expected due to the quenching of magnon excitations. 

\begin{figure}[!htb]
\centering
\includegraphics[width=0.9\linewidth]{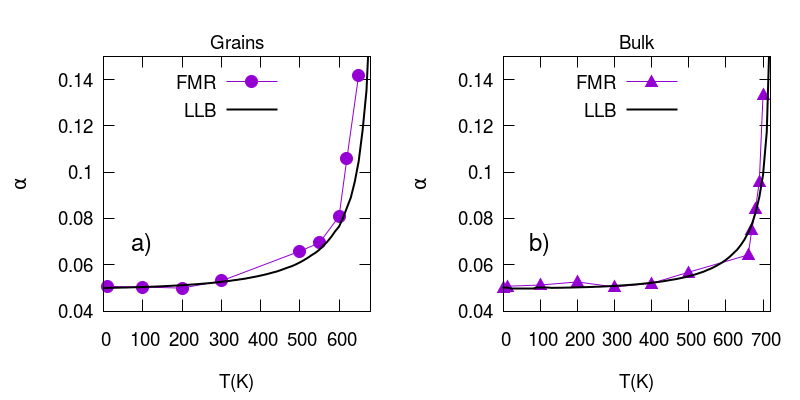}
\caption{ Damping as function of temperature for granular (5nm $\times$ 5nm $\times$ 10nm) (a) and bulk (b) systems. The damping calculated via FMR method is compared against the effective damping from parameterised the LLB formalism - Eq.~9, where $m(T,D)$ and $T_c(D)$ are computed numerically from the atomistic model.
\label{fmr_grains_bulk_llb}}
\end{figure}

\begin{figure*}[!htb]
\includegraphics[trim={0 0 0 2cm},width=0.9\linewidth]{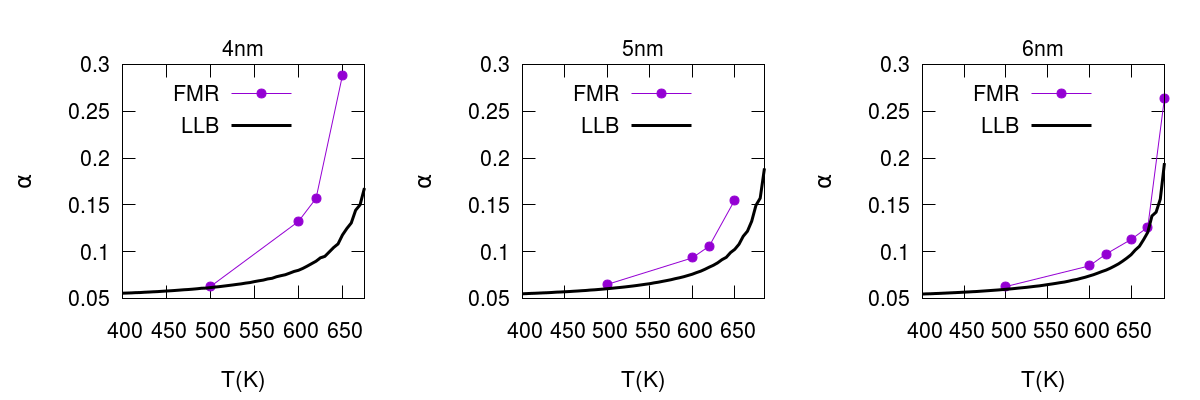}
\caption{Temperature dependence of the damping constant for diameters of 4nm, 5nm and 6nm. Solid lines are calculations using the LLB damping expression. Divergence from the LLB expression for small particle diameter is indicative of surface effects.
\label{fmr_cube_size}}
\end{figure*}

With increasing temperature, for both bulk and granular systems, the effective damping  increases. This can be understood as, with enhanced temperature, there is increasing excitation of magnons which can suffer more complex non-linear scattering processes.

Moving on to granular systems, the inclusion of the surface will add extra magnon modes into the system, leading to more scattering effects that will increase the effective damping. In order to effect a qualitative illustration of surface effects we use the damping calculated from the Landau-Lifshitz-Bloch equation ~\cite{garanin1997fokker}. An analytical solution to the variation of damping with temperature exists in the LLB description, as given by Garanin \cite{garanin1997fokker} and Ostler \textit{et al} \cite{Ostler2014}. The effective damping as derived within the LLB description is given by:
\begin{equation}\label{damping_llb}
    \alpha (T) = \frac {\lambda}{m(T)} \left (1- \frac{T}{3T_c} \right ) ,
\end{equation}
where $\lambda$ is the input coupling to the thermal bath used in atomistic spin dynamics simulations, $\Tc$ the Curie Temperature of the system, $m(T)= M(T)/M_sV$ the normalised magnetisation. In principle Eq.~\ref{damping_llb} is strictly valid only for an infinite system. However, as a first approximation, finite size effects can be introduced empirically using diameter dependent functions $m(T,D)$ and $T_c(D)$ calculated using an atomistic model. In the damping calculations considered here, the grain surfaces have two effects. Firstly, the loss of coordination at the surfaces drive a reduction in $T_c$ and loss of criticality of the phase transition. This effect can be accounted for by using numerically calculated $m(T,D)$ and $T_c(D)$ for a given diameter D. The second effect is the increased magnon scattering at the surfaces which is a dynamic effect and not included in the parameterization of the static properties.  Thus it seems reasonable to associate deviations from the parameterized version of Eq.~\ref{damping_llb} with scattering at the grain surfaces.

Consequently, we compare our numerical results for $\alpha (T,D)$ with the parameterised version of Eq. \ref{damping_llb} - Fig. \ref{fmr_grains_bulk_llb} , where $m(T,D)$ is calculated numerically with the ASD model (shown in Fig.~ \ref{fmr_damping_grain}, inset) and the Curie Temperature ($T_c(D)$) is calculated from the peak of the susceptibility. For the bulk system, the numerical damping calculated from the FMR curves with the atomistic model agree well with the damping calculated with the analytical formula given by Eq.~\ref{damping_llb}. This is consistent with the first comparison of atomistic and LLB models \cite{chubykalo2006dynamic} which showed that the mean-field treatment of~\cite{garanin1997fokker} agreed quantitatively well with atomistic model calculations for the transverse and longitudinal damping. However, the granular system gives a consistently increased damping compared to the analytical formula. Following the earlier reasoning, this enhancement can be attributed to the scattering effects at the grain surface.

To systematically study the effect of the scattering at the surface, we have calculated the damping as a function of the system size. For simplicity, we consider cubic grains with a volume varying from 4nm $\times$ 4nm $\times$ 4nm, 5nm $\times$ 5nm $\times$ 5nm and 6nm $\times$ 6nm $\times$ 6nm. Fig.~\ref{fmr_cube_size} shows the damping as a function of the temperature for the different system sizes. With decreasing grain size, the damping is enhanced, and systematically diverges further from the LLB analytical damping. The separation of the effects of the finite size on the static and dynamic properties through comparison with the parameterised version of Eq.~\ref{damping_llb} strongly suggests that this is due to surface scattering of magnons. Clearly, the magnon contributions to the damping give rise to an increase of the damping with increasing temperature, which is inconsistent with the results of Richardson et al.~\cite{Richardson2018}. However, the experiments described in ref~\cite{Richardson2018} give the temperature dependence of the linewidth which likely has contributions from inhomogeneous line broadening arising from dispersion of magnetic properties. In the following we develop a model accounting for the inhomogenous line broadening which gives good qualitative agreement with the experiments.

\section{Model including inhomogeneous line broadening}

\begin{figure*}[!htb]
\centering
\includegraphics[trim={0 0 0 0cm},width=0.85\linewidth]{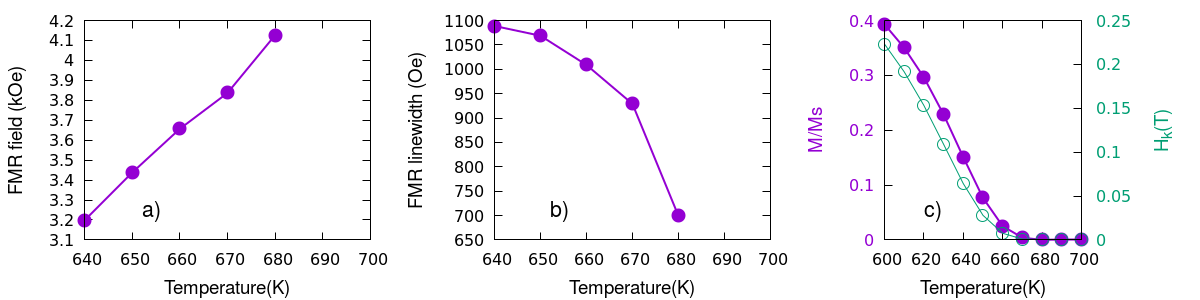}
\caption{Field swept FMR for a lognormal distribution of grains of $D=4$ nm and $\sigma_D=0.17$. Input damping $\lambda=0.01$, input $H_k^0=0.66$T, $f=13.7$ GHz. The anisotropy is lower than for bulk FePt to (a) allow resonance at a frequency of 13.7Ghz corresponding to experiment. The figure shows (a) the variation of the FMR field, (b) FMR linewidth ($\Delta H$) and (c) system magnetisation and anisotropy field as function of temperature. Close to $\Tc$, the linewidth shows a decrease which translates to a decrease in the damping of the system. No magnetostatic or exchange interaction between grains is considered.} 
\label{decreasing_lw}
\end{figure*}

Realistic granular systems will present a distribution of properties. In the simplest case, the distribution of magnetic properties can arise from a distribution of the size of the grains, which can induce a distribution of $\Tc$, $m$ and $H_k$. Since  it is computationally expensive to study a system of grains numerically within the ASD model we can, in the first instance, model the effect of the distributions analytically. In the case of a distribution of grains of diameter $D$, the power spectrum of the system is expressed by: 
\begin{equation}\label{distP}
   P^{sys}(f,T)=\int_0^{+\infty} P(f,D,T) F(D) d D
\end{equation}

The distribution of size, $F(D)$, is considered lognormal. The power spectrum of a grain of diameter $D$ can be expressed by \cite{Ostler2014}:
\begin{equation}\label{power}
   P(f,B_0,D,T)=C \Tilde{m} D^2 \frac {f^2 \gamma B_0^2  \Tilde{\alpha}}{(\Tilde{\alpha}\Tilde{f_0})^2+(f-\Tilde{f_0})^2} 
\end{equation}

where $\Tilde{m}=m(T,D), \Tilde{\alpha}=\alpha(T,D), \Tilde{f_0}=f_0(T,D)=\gamma (B_0+Hk(T,D)), C=\frac{\pi h}{16}$. This allows to model both frequency swept FMR ($B_0=$constant) and field swept FMR ($f$=constant). 

We note that a Distribution of grain size leads to  distributions of further properties, starting, due to finite size effects, with the Curie temperature. Each of these is introduced into the analytical model as follows. Hovorka et al \cite{hovorka2012curie} have shown via finite size scaling analysis that the relation between the size of a grain and its Curie temperature is given by:
\begin{equation}\label{var_tc}
    \Tc(D)=\Tc^{\infty }(1 - d_0/D)^{1/\nu} ,
\end{equation}
where $d_0=0.71$ and $\nu=0.79$ \cite{hovorka2012curie} parametrised for nearest neighbours exchange systems and $\Tc^{\infty}=720K$. The variation in $\Tc$ will introduce a variation in the magnetisation curves given by:
\begin{equation}\label{var_ms}
    m(T,D)=\left(1-\frac{T}{\Tc(D)}\right)^\beta, \beta=0.33.
\end{equation}
As a further consequence, the anisotropy will be dependent on the diameter. The uniaxial anisotropy energy $K$ has a temperature dependence in the form of $K(T)\sim m(T)^\gamma $. For FePt it was found that the exponent is equal to 2.1 by experimental measurements \cite{thiele2002temperature}\cite{okamoto2002chemical} in agreement with later \textit{ab-initio} calculations \cite{mryasov2005temperature} . Hence the anisotropy field will depend on  $m(T,D)$ with an exponent of 1.1;
\begin{equation}\label{var_hk}
    H_K(T,D)=H_K^0 m(T,D)^{1.1},
\end{equation}
leading to a dispersion of $H_K(D,T)$.

Finally the size distribution will produce  different variations of damping  as a function of grain size, since 
\begin{equation}\label{var_alp}
    \alpha(T,D)=\frac {\lambda} {m(T,D)} \left(1-\frac{T}{3T_c(D)}\right)\mathrm{.}
\end{equation}

In the presence of distributions of properties, the variation of damping with temperature can have a complex behaviour, especially close to $T_c$ where there is a strong variation of magnetic properties with temperature and size. On reverting to a monodispersed system by setting the distributions to $\delta$-functions the damping is given by Eq.~\ref{var_alp} resulting in an increase of linewidth consistent with temperature Fig.~\ref{fmr_cube_size}.

Richardson \textit{et al} \cite{Richardson2018} have shown that, in the case of a granular system of FePt, close to \Tc a decrease in damping/linewidth is observed. This effect was attributed to the competition between two-magnon scattering and spin-flip magnon electron scattering. We have shown via atomistic spin dynamics simulations that  surface effects alone cannot be responsible for a decrease in damping, scattering at the surface leading to increased damping at high temperatures.

It is well known that, in the presence of a distribution of properties in the system, the linewidth broadens. In our case the distribution of size will lead to a distribution of anisotropy which increases the linewidth. Close to the Curie Temperature of the system, some grains will become superparamagnetic and will not contribute further to the FMR spectrum, hence it is possible that close to $T_c$, the linewidth can decrease. Fig. \ref{decreasing_lw} presents a case where a decrease in linewidth appears within 30-40K of the Curie Temperature of the system, a similar temperature interval as spanned by the experimental measurements \cite{Richardson2018}. Fig. \ref{decreasing_lw} a) and b) show the variation of the FMR field and linewidth as functions of temperature. The FMR spectra are calculated at constant frequency of $f=13.7GHz$, consistent with the experimental value used in \cite{Richardson2018}. The average magnetisation of distributed grains is calculated as 
\begin{equation}
    M(T)=\frac{\int_0^\infty m(T,D)F(D) D^2 dD}{\int_0^\infty F(D) D^2 dD} 
\end{equation} 
and the anisotropy field is calculated as 
\begin{equation}
H_K(T)=\int_0^\infty H_K(T,D)F(D)dD\mathrm{.} 
\end{equation} 
The decrease in linewidth is associated with the fact that, close to the Curie Temperature of the system, the small grains become superparamagnetic and do not contribute to the power spectrum.  The loss of the signal from the small grains is especially pronounced due to the enhanced damping of smaller grains.

\section{Conclusions and outlook}
We have calculated the temperature dependence of damping and anisotropy for small FePt grain sizes. These parameters were calculated  within the ASD framework, via simulation of swept frequency FMR processes and a fitting procedure based on the grid search method. The grid search method offers a much faster determination of damping and anisotropy, parameters crucial for the development of future generation HAMR drives. The method can be applied both for numerical data, as well for experimental relaxation curves obtained via pump- probe experiments. The damping calculations at large temperatures showed an increased damping for uncoupled granular systems as expected due to increased magnon excitation at high temperature. Deviations from the parameterised expression for the temperature dependence of damping from the LLB equation with decreasing grain size suggest that scattering events at  grain boundaries enhance the damping mechanism.

This increase in damping, however, is not consistent with the experimental data of Richardson~\textit{et al}~\cite{Richardson2018} which show a decrease in linewidth at elevated temperatures. We have developed a model taking into account inhomogeneous line broadening arising from the size distribution of the grains which gives rise to  concomitant dispersions of $\Tc$, $m$ and $K$. The model has been used to simulate swept field FMR as used in the experiments.  Calculations have shown that, under the effect of distribution of properties, the linewidth can exhibit a decrease towards large temperatures, in  accordance with the experiments of Ref.~\cite{Richardson2018}. The decrease is predominantly due to a transition to superparamagnetic behaviour of small grains with increasing temperature. This suggests  inhomogeneous line broadening (likely a significant factor in granular films) as an explanation for the unusual decrease in linewidth measured by Richardson \textit{et al} \cite{Richardson2018}. As large damping is necessary for good performance of HAMR and MRAM devices with this work we further stress the importance of experimentally controlling the size distributions of the media. 

\section{Acknowledgements}
We are grateful to Prof. M Wu and Stuart Cavill for helpful discussions. Financial support of the 
Advanced Storage Research Consortium   is gratefully acknowledged.
The  atomistic  simulations  were  undertaken on the VIKING cluster, which is a high performance compute  facility  provided  by  the  University  of  York.    We are grateful for computational support from the University of York High Performance Computing service,VIKING and the Research Computing team.

\end{document}